\documentclass[11pt,a4paper]{article}

\usepackage{graphicx}
\usepackage{color}
\usepackage{amsmath,amssymb}
\usepackage{fdsymbol}
\usepackage{bbold}
\usepackage{t1enc}

\usepackage{color}
\usepackage[dvipsnames]{xcolor}

\DeclareMathSymbol{\shm}{\mathbin}{AMSa}{"39}
\newcommand{\oh}{\frac{1}{2}}

\newcommand{\UU}{|{ \small ++} \rangle}
\newcommand{\UD}{|{ \small +-} \rangle}

\newcommand{\DU}{|{ \small -+} \rangle}
\newcommand{\DD}{|{ \small --} \rangle}

\begin{document}

\begin{center}
\begin{Large}
{\bf Entanglement autodistillation from particle decays}
\end{Large}

\vspace{0.5cm}
\renewcommand*{\thefootnote}{\fnsymbol{footnote}}
\setcounter{footnote}{0}
J.~A.~Aguilar-Saavedra, J.~A. Casas \\[1mm]
\begin{small}
Instituto de F\'isica Te\'orica IFT-UAM/CSIC, c/Nicol\'as Cabrera 13-15, 28049 Madrid, Spain \\
\end{small}
\end{center}

\begin{abstract}
Particle decays do not constitute a spin ``measurement'' in the quantum-mechanical sense, but still modify the spin state, in particular for an entangled system. We show that for a spin-entangled pair of particles the entanglement of the system can increase after the decay of one particle. This unique phenomenon has no equivalent for stable particles and could be observable in top pair production at a high-energy polarized $e^+ e^-$ collider.
\end{abstract}

\section{Introduction}

Quantum entanglement stands out as one of the most intriguing phenomena in the realm of quantum mechanics~\cite{Einstein:1935rr,sch1935}. It has been experimentally verified thoroughly, and has profound implications for the development of quantum technologies, quantum computing, and for our philosophical conception of the universe. Even though tests and applications of quantum entanglement date from decades ago, elementary particle physics provides novel effects, still untested, which involve quantum entanglement and particle decay: post-decay entanglement~\cite{Aguilar-Saavedra:2023hss} and post-selection of decayed states~\cite{Aguilar-Saavedra:2023lwb} (see also~\cite{Bernabeu:2019gjs}). In this Letter we show a remarkable feature: the possibility that the state after decay is {\em more} entangled than before.

Any reasonable measure of entanglement requires monotonicity, i.e. the entanglement cannot increase under local operations and classical communications (LOCCs).
Still, under suitable local manipulations the entanglement may increase with some probability of failure. These generalized operations are called SLOCCs (i.e. stochastic LOCCs) \cite{Bennett:2000fte}. This is the basis of the entanglement distillation techniques. We will show that, for a system of two particles $A$ (Alice) and $B$ (Bob), when one of them decays the system can   spontaneously get more entangled (under any standard measure of entanglement), even though, obviously, the process is completely local. We will see that this entanglement amplification --- which can be dubbed as ``autodistillation'' --- would be experimentally observable in top pair production at a high-energy $e^+ e^-$ collider, especially if the beams are polarized. Top quarks are ideally suited to observe this effect. 
Namely, they decay $t \to W b$ with a branching ratio near unity and the spin observables of both $t$ and $W$, as well as their spin correlations, can be measured from angular distributions of their decay products~\cite{Kane:1991bg,Aguilar-Saavedra:2015yza,Rahaman:2021fcz,Ashby-Pickering:2022umy,Bernal:2023jba}. 
In this way, spin entanglement of $t \bar t$ pairs 
at the Large Hadron Collider (LHC) has already been established~\cite{ATLAS:2023fsd,CMS:2024pts}. 
In $e^+ e^-$ collisions, and depending on the kinematical configuration of the decay products, the entanglement of $t$ and $W^-$ (produced in the decay of $\bar t$) or, equivalently, $\bar t$ and $W^+$, can be larger than the initial one.


\section{Formalism of post-decay density operators}

Let us consider a particle $A$ in a pure state $|\psi\rangle$, decaying into some given final state, $A \to A_1 \dots A_n$ with defined  four-momenta. This can be viewed as a two-step process: (1) the decay of $|\psi\rangle$ into a state $T|\psi\rangle$, with $T$ the transition operator. This accounts for the multi-particle component of the unitary evolution of the initial state.
(2) the {\em measurement} of the momenta of $A_1, \dots,  A_n$,  which we will generically label as $P$. This measurement implicitly involves the identification of the various particle species.
Then, we determine the final state by applying to $T |\psi\rangle$ the projector into the subspace with a definite value of the final momenta,
\begin{equation}
\mathcal{P} = \sum_\alpha |P  \, \xi_\alpha \rangle \langle P \, \xi_\alpha | \,,
\end{equation}
with $\xi_\alpha$ generically denoting spin indices of the multi-particle final state. 
 The final state is then
\begin{equation}
|\psi^\prime \rangle = \mathcal{N} 
\sum_\alpha |P\, \xi_\alpha\rangle \langle P \, \xi_\alpha | T | \psi \rangle  = \mathcal{N} |P\rangle \sum_\alpha \langle P \, \xi_\alpha | T | \psi \rangle |\xi_\alpha\rangle
\end{equation}
The normalisation can be conveniently chosen so that
\begin{equation}
|\mathcal{N}|^2 \sum_\alpha |\langle P \, \xi_\alpha | T | \psi \rangle |^2 = 1 \,.
\end{equation}
Writing the initial state  in a basis of spin eigenstates, $
|\psi\rangle = \sum_j c_j |\phi_j \rangle$, and defining the decay amplitudes
\begin{equation}
 M_{\alpha j} \equiv \langle P \, \xi_\alpha | T | \phi_j \rangle   
\end{equation}
the normalized final state reads
\begin{equation}
|\psi^\prime\rangle = \frac{1}{\|Mc\, \|} |P \rangle \sum_\alpha (Mc)_\alpha |\xi_\alpha \rangle \,.
\end{equation}
where $Mc$ is a vector with components $(Mc)_\alpha = \sum_j M_{\alpha j} c_j$, and $\| \cdot \|$ is the Euclidean norm.

For a mixed state characterised by the operator $\rho = \sum_{ij} \rho_{ij} |\phi_i \rangle \langle \phi_j |$
the final state is also obtained by considering the decay as the two-step process of decay plus measurement. The resulting density operator is
\begin{equation}
\rho' = \frac{1}{\operatorname{tr} (\mathcal{P} \tilde \rho \, \mathcal{P})}
\mathcal{P} \tilde \rho \, \mathcal{P} \,,
\end{equation}
with $\tilde \rho = T \rho \, T^\dagger$.
In terms of the decay amplitudes $M_{\alpha j}$, and ignoring the spatial part $|P\rangle$ which is fixed by the measurement, the spin density operator can be written as
\begin{equation}
\rho' = \frac{1}{\sum_i (M \rho M^\dagger)_{\alpha\alpha}}
\sum_{\alpha\beta} (M \rho M^\dagger)_{\alpha\beta} |\xi_\alpha \rangle \langle \xi_\beta | \,.
\end{equation}
Above, matrix multiplication is understood, that is, $(M \rho M^\dagger)_{\alpha\beta} = M_{\alpha i} \rho_{ij} M^\dagger_{j\beta}$.

This formalism can easily be generalized to two or more particles, by restricting the projector to the subspace of the decaying particle. 
Let us assume we have a system of two particles $A$, $B$ in a pure state
\begin{equation}
|\psi \rangle = \sum_{ij} c_{ij} | \phi_i \chi_j \rangle \,,
\label{initial}
\end{equation}
where $A$ is the decaying particle as before, and $|\chi_j\rangle$ describe the spin degrees of freedom of particle $B$, whose possible decay we do not consider. The state after the decay is, omitting the spatial part $|P \rangle$,
\begin{equation}
|\psi' \rangle = \frac{1}{\|Mc\, \|} \sum_{\alpha j} (Mc)_{\alpha j} |\xi_\alpha \chi_j \rangle \,,
\label{final}
\end{equation}
with $(Mc)_{\alpha j} = M_{\alpha k} c_{kj}$. The norm of the matrix $Mc$ is the usual one $\| A \|^2 = \operatorname{tr} A A^\dagger$. In the most general case, the two-particle spin state is given by a density operator
\begin{equation}
\rho = \sum_{ijkl} \rho_{ij}^{kl} |\phi_i \chi_k \rangle \langle \phi_j \chi_l | \,.
\end{equation}
 The spin state after the decay of $A$ is given by 
\begin{equation}
\rho' = \frac{1}{\sum_{\alpha k} (M \rho^{kk} M^\dagger)_{\alpha\alpha}} 
\sum_{\alpha\beta kl} (M \rho^{kl} M^\dagger)_{\alpha\beta} |\xi_\alpha \chi_k \rangle \langle \xi_\beta \chi_l | \,,
\label{ec:rhop}
\end{equation}
with matrix multiplication in the lower indices of $\rho_{ij}^{kl}$.

Measurements of $\rho'$ at fixed values of $|P\rangle$ are not possible and, in practice, they have to be replaced by measurements over a region $S$ in the phase space for $A$ decay. The most common situation is a two-body decay, in which the phase space is two-dimensional and can be parameterised by two angles $\Omega=(\theta,\phi)$. The post-decay density operator for this case is 
\begin{equation}
\rho' = \frac{1}{\sum_{\alpha k} \int_S d\Omega \, (M \rho^{kk} M^\dagger)_{\alpha\alpha}}   \sum_{\alpha\beta kl} \left[ \int_S d\Omega \, (M \rho^{kl} M^\dagger)_{\alpha\beta} \right] |\xi_\alpha \chi_k \rangle \langle \xi_\beta \chi_l | \,.
\label{ec:rhopint}
\end{equation}
The generalization of the above equation for multi-body decay phase space is straightforward. We note that the density operator $\rho$ may already involve integration in phase space, but this does not affect the calculation of $\rho'$ using either (\ref{ec:rhop}) or (\ref{ec:rhopint}). This formalism of post-decay density operators can be compared with direct Monte Carlo calculations, finding excellent agreement~\cite{Aguilar-Saavedra:2024hwd}.

\section{Understanding autodistillation}

With the previous formalism we can show how the state after decay can be {\em more} entangled than before, using a simple example. Let us consider a system composed of two particles $A$ and $B$ in a pure state, namely Eq.~(\ref{initial}), described by the matrix $c$. 
A convenient measure of the entanglement of a pure state is the concurrence, defined as 
$C^2=2 (1-{\rm Tr}\ \rho_A^2)$, where $\rho_A$ is the effective density matrix in Alice side after tracing in Bob (or the other way around).
Hence, the initial concurrence is 
\begin{equation}
C^2_\text{initial} = 2 \left(1 - \operatorname{tr} [c c^\dagger]^2 \right)\,.
\end{equation}
The state after the decay is still pure, given by (\ref{final}), with a concurrence
\begin{equation}
C^2_\text{final} = 2 \left( 1 - \frac{\operatorname{tr} [(Mc) (Mc)^\dagger]^2}{(\operatorname{tr} [(Mc) (Mc)^\dagger])^2} \right)\,.
\end{equation}
Now, depending on the matrix elements, $M_{\alpha j}$, this entanglement can be higher than the initial one.

Let us consider two spin-1/2 particles, say $a, \bar a$, with well-defined momenta, in a spin-entangled state
\begin{equation}
|\psi \rangle = c_\alpha |a_{\frac{1}{2}} \bar a_{-\frac{1}{2}}\rangle + s_\alpha |a_{-\frac{1}{2}} \bar a_{\frac{1}{2}} \rangle 
\label{initial_example}\,,
 \end{equation}
where $c_\alpha =\cos \alpha$, $s_\alpha = \sin\alpha$ and the subscripts $\pm 1/2$ indicate the eigenvalue of the third spin component.
This equation corresponds to the initial state $|\psi \rangle$ in Eq.~(\ref{initial}) with $c_{\frac{1}{2}, -\frac{1}{2}}=c_\alpha$, $c_{-\frac{1}{2}, \frac{1}{2}}=s_\alpha$. In this case the reduced density matrix reads $\rho_A={\rm diag}\{c_\alpha^2,\ s_\alpha^2\}$, and thus the initial concurrence is
\begin{equation}
C^2_{\rm initial}=2\left(1-(c_\alpha^4+s_\alpha^4)\right)= 4c_\alpha^2s_\alpha^2\,.
\label{conc_initial}
 \end{equation}
Now suppose that the particle $a$ decays along the process 
$a \to b  V$, where $b$ is a spin 1/2 particle with tiny mass which we neglect here and well-defined chirality (and thus helicity), say left-handed, and $V$ is a massive vector boson. This is the case for the top decay, $t\rightarrow b_L W$, but for the moment we prefer not to particularize to specific particles. Suppose that in the rest frame of the $a-$particle the decay has occurred with the momentum of the vector boson along the (positive) $z-$direction. The precise meaning of this statement is the following: in the Alice side (particle $a$) we detect a particle $b$, originated by the decay, and measure its momentum, which instantaneously determines the momentum of $V$. It is irrelevant whether or not we measure the spin of $b$ since this is established by its left-handed chirality and its negligible mass. Actually, in this instance the spin state of $V$ is also completely determined by that of $a$ and angular momentum conservation. The only two possibilities are
\begin{equation}
a_{\frac{1}{2}}\ \longrightarrow\ b_{-\frac{1}{2}}\ V_{0},\ \ \ \ a_{-\frac{1}{2}}\ \longrightarrow\ b_{-\frac{1}{2}}\ V_{-1} \,,
\end{equation}
where the subscripts of $b,V$ denote their helicities. The amplitudes for these processes can be labelled as $M_{\frac{1}{2}}$ and $M_{-\frac{1}{2}}$.
Hence, the final state after the decay, Eq.~(\ref{final}) reads
\begin{equation}
|\psi' \rangle = {\cal N}|P\rangle \left[ c_{\alpha} 
M_{\frac{1}{2}}
\,  |b_{-\frac{1}{2}}V_0 \, \bar a_{-\frac{1}{2}} \rangle + s_{\alpha}M_{-\frac{1}{2}}\,  |b_{-\frac{1}{2}}V_1 \, \bar a_{\frac{1}{2}} \rangle\right] \,.
\label{almost_final_example}
\end{equation}
Since the spin state of $b$ factorizes in (\ref{almost_final_example}), we can write the final state as
\begin{equation}
|\psi' \rangle = {\cal N}|P\rangle|b_{-\frac{1}{2}}\rangle \left[c_{\alpha}M_{\frac{1}{2}} \, |V_0 \,  \bar a_{-\frac{1}{2}} \rangle + s_{\alpha}M_{-\frac{1}{2}} \,  |V_1 \,  \bar a_{\frac{1}{2}} \rangle\right] \,.
\label{final_example}
\end{equation}
Comparing Eqs.~(\ref{initial_example}) and (\ref{final_example}) we observe that the initial entangled $\{a,\bar a\}$ system has led, after the decay of $a$, to an entangled $\{V,\bar a\}$ system with different amount of entanglement. The concurrence for the final spin state is
\begin{equation}
C^2_{\rm final}= 4c_\alpha^2s_\alpha^2\left(
c_\alpha^2 r + s_\alpha^2 r^{-1}
\right)^{-2} \,,
\label{conc_final}
 \end{equation}
with $r=\left| M_{\frac{1}{2}} \;/\; 
M_{-\frac{1}{2}}\right|$.
The point is that if $r\neq 1$, which is the usual case, then  $C^2_{\rm initial}\neq C^2_{\rm final}$. For example, if the system is initially prepared in a state (\ref{initial_example}) with $\tan \alpha= r$, we get a final state with maximum entanglement, $C^2_{\rm final}=1$.


\section{Top polarized decay amplitudes}

In order to study autodistillation in a realistic setup we consider top quarks and their decay $t \to Wb$. In the top quark rest frame we take a reference system $(x,y,z)$, and denote by $\vec p = q (\sin \theta \cos \phi,\sin \theta \sin \phi,\cos \theta)$ the three-momentum of the $W$ boson, and $E_W$ and $E_b$ the energies of $W$ and $b$, all quantities evaluated in the top quark rest frame. The masses of $t$, $W$ and $b$ are denoted as $m_t$, $M_W$ and $m_b$, as usual. Unless otherwise indicated, we take $m_t = 172.5$ GeV, $M_W = 80.4$ GeV, $m_b = 4.8$ GeV, and consider a generic interaction
\begin{equation}
\mathcal{L} = - \frac{1}{\sqrt 2} \bar b \gamma^\mu (g_L P_L + g_R P_R) t \, W_\mu^-  \,.
\label{ec:lagr}
\end{equation}
The standard model interaction is recovered by setting $g_L = g$, $g_R = 0$.

The decay amplitudes $M_{\alpha j}$ appearing in a general post-decay density operator (\ref{ec:rhopint}) must be calculated for a fixed reference system. However, for fixed angles $(\theta,\phi)$ the helicity amplitudes can be used as well, as they are related by a change of basis for the $b$ and $W$ spins. Helicity amplitudes for the decay of top quarks have been provided in previous literature~\cite{Boudjema:2009fz}. We calculate them here, also including those for anti-quarks.
In the helicity basis the top spin is quantized along the $\hat z$ axis, whereas for the $W$ boson and $b$ quark the helicity direction is used. We label the amplitudes as $A_{s_1\,s_2\,s_3}$, where $s_1$, $s_2$, $s_3$ are the spin components of $t$, $b$ and $W$, in the respective axis considered. For these amplitudes we introduce the kinematical factors
\begin{align}
& H_1^\pm = [m_t (E_b+m_b)]^{1/2} \left( 1 \pm \frac{q}{E_b+m_b} \right) \,, \notag \\
& H_2^\pm = \frac{E_W \pm q}{\sqrt{2} M_W} \,.
\end{align}
Notice that $H_1^-$ vanishes for $m_b = 0$, and the terms with this factor are helicity-suppressed. The non-zero amplitudes are then
\begin{align}
& A_{\oh\,\oh\,1} = \frac{1}{\sqrt{2}} \left( g_L H_1^- - g_R H_1^+ \right) e^{i \phi/2} \cos \frac{\theta}{2} 
\,, \notag \\
& A_{\oh\,\oh\,0} = - \frac{1}{\sqrt 2} \left( g_L H_1^- H_2^- - g_R H_1^+ H_2^+ \right) e^{i \phi/2} \sin \frac{\theta}{2}  
\,, \notag \\
%
& A_{\oh\,-\!\oh\,0} =  \frac{1}{\sqrt 2} \left( g_L H_1^+ H_2^+ - g_R H_1^- H_2^- \right) e^{i \phi/2} \cos \frac{\theta}{2} 
\,, \notag \\
& A_{\oh\,-\!\oh\, -\!1} = - \frac{1}{\sqrt{2}} \left( g_L H_1^+ - g_R H_1^- \right) e^{i \phi/2} \sin \frac{\theta}{2}  
\,, \notag \\
& A_{-\!\oh\,\oh\,1} = \frac{1}{\sqrt 2} \left( g_L H_1^- - g_R H_1^+ \right) e^{-i \phi/2}  \sin \frac{\theta}{2} \,, \notag \\
& A_{-\!\oh\,\oh\,0} = \frac{1}{\sqrt 2} \left( g_L H_1^- H_2^- - g_R H_1^+ H_2^+ \right)  e^{-i \phi/2} \cos \frac{\theta}{2} \,, \notag \\
& A_{-\oh\,-\!\oh\,0} = \frac{1}{\sqrt 2} \left( g_L H_1^+ H_2^+ - g_R H_1^- H_2^- \right)  e^{-i \phi/2} \sin \frac{\theta}{2} \,, \notag \\
& A_{-\oh\,-\!\oh\,-1} = \frac{1}{\sqrt 2} \left( g_L H_1^+ - g_R H_1^- \right) e^{-i \phi/2} \cos \frac{\theta}{2} \,,
\label{ec:amphel}
\end{align}
with $A_{s_1 \oh\,-\!1} = A_{s_1 -\!\oh\,1} = 0$ due to angular momentum conservation. They agree with those in Ref.~\cite{Boudjema:2009fz}. The amplitudes $\bar A$ for top anti-quark decays have also been obtained. They relate to those in (\ref{ec:amphel}) by the interchange of left- and right-handed couplings, namely $\bar A_{s_1\,s_2\,s_3}(\theta,\phi) = \left. A_{s_1\,s_2\,s_3}(\theta,\phi) \right|_{g_L \leftrightarrow g_R}$.
Note that, neglecting the bottom mass, for $\theta=0,\pi$ only two amplitudes survive, which essentially correspond to the $M_{\frac{1}{2}}, M_{-\frac{1}{2}}$ amplitudes in Eq.~(\ref{almost_final_example}). Depending on the value of $\theta$, the value of $r$ in Eq.~(\ref{conc_final}) is larger or smaller than $1$, which reflects the probabilistic character of the entanglement amplification.


\section{Autodistillation in $e^+ e^-$ collisions}

Quantum entanglement between top quark pairs produced at the LHC has been addressed in in previous literature~\cite{Afik:2020onf,Fabbrichesi:2021npl,Severi:2021cnj,Afik:2022kwm,Aguilar-Saavedra:2022uye,Afik:2022dgh,Dong:2023xiw,Han:2023fci}.
However, top quark pair production at the LHC is not suited to observe an entanglement increase. Near threshold, the $t \bar t$ density operator is dominated by the spin-singlet component $1/\sqrt{2} \left[ \UD - \DU \right]$, and in the boosted central region by the triplet $1/\sqrt{2} \left[ \UU + \DD \right]$. But, obviously, from such initial states (i.e. with $c_\alpha = s_\alpha$) the entanglement cannot increase, see Eqs.~(\ref{initial_example}) and (\ref{final_example}).
In order to see the reason behind that, we write the $t \bar t$ density operator in terms of polarizations $B_i^+$ (for the quark), $B_i^-$ (for the anti-quark) and spin correlation coefficients $C_{ij}$,
\begin{equation}
\rho = \frac{1}{4}\left(
\mathbb{1}\otimes \mathbb{1} +\sum_i(B_i^+ \sigma_i\otimes \mathbb{1} + B_i^- \mathbb{1}\otimes \sigma_i)  + \sum_{ij}C_{ij} \sigma_i\otimes \sigma_j
\right) \,.
\label{generalrho}
\end{equation}
In the basis of $S_3$ eigenstates
$\{  \UU \,, \UD \,, \DU  \,, \DD  \}$
the density operator (\ref{generalrho}) reads
\begin{tiny}
\begin{equation*}
\rho= \frac{1}{4}\left[\begin{array}{cccc} 
1+B_3^++B_3^-+C_{33} & B_1^-+C_{31}-i(B_2^-+C_{32})& B_1^++C_{13}-i(B_2^++C_{23})&
C_{11}-C_{22}-i(C_{12}+C_{21})\cr
B_1^-+C_{31}+i(B_2^-+C_{32})&
1+B_3^+-B_3^--C_{33}&
C_{11}+C_{22}+i(C_{12}-C_{21})&
B_1^+-C_{13}-i(B_2^+-C_{23})\cr
B_1^++C_{13}+i(B_2^++C_{23})&
C_{11}+C_{22}-i(C_{12}-C_{21})&
1-B_3^++B_3^--C_{33}&
B_1^--C_{31}-i(B_2^--C_{32})\cr
C_{11}-C_{22}+i(C_{12}+C_{21})&
B_1^+-C_{13}+i(B_2^+-C_{23})&
B_1^--C_{31}+i(B_2^--C_{32})&
1-B_3^+-B_3^-+C_{33}
\end{array}\right]
\label{rhoexpl}
\end{equation*}
\end{tiny}
Then, one can see that the reason for having eigenvectors with $c_\alpha = s_\alpha$ is the fact that $t$ and $\bar t$ polarizations vanish at the leading order (with a tiny polarization generated at higher orders), as off-diagonal spin correlations $C_{13}$, $C_{23}$ do. Depending on the reference system $(x,y,z)$ used, a small $C_{12}$ may be present but it only breaks the equality $c_\alpha = s_\alpha$ in the principal eigenvector at the permille level. 

\begin{figure}[htb]
\begin{center}
\includegraphics[height=6.5cm,clip=]{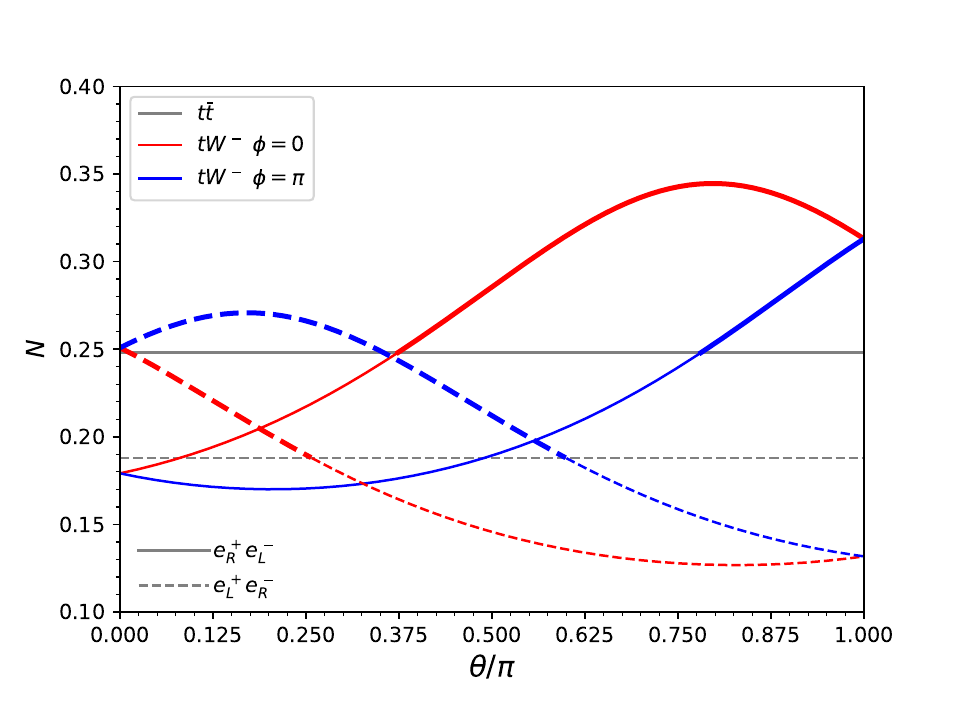} 
\caption{$tW^-$ entanglement measure $N$ as a function of $\theta$, for $e_R^+ e_L^-$ (solid lines) and $e_L^+ e_R^-$ (dashed lines) collisions at 1 TeV with $|\cos \theta_t| \leq 0.2$. The horizontal lines indicate the $t \bar t$ entanglement.}
\label{fig:Nvstheta}
\end{center}
\end{figure}

In order to have polarized top quarks, one has to resort to $e^+ e^-$ collisions, where the chirality of the $ttZ$ coupling produces non-zero $B_i^\pm$. However, top pairs produced in $e^+ e^-$ collisions near threshold have their spins in a separable state: For an initial $e_L^+ e_R^-$ state, both electron spins point in the electron direction and, by  angular momentum conservation, so do the $t$ and $\bar t$ spins. For an initial $e_R^+ e_L^-$, it is the opposite. Therefore, in order to have entanglement, we need to consider energetic collisions, in which case the helicity basis is a convenient choice, with axes
$\hat z = \hat k$, $\hat x = \hat r$, $\hat y = \hat n$, the K, R and N axes being defined as~\cite{Bernreuther:2015yna}
\begin{itemize}
\item K-axis (helicity): $\hat k$ is a normalised vector in the direction of the top quark three-momentum in the $t \bar t$ rest frame.
\item R-axis: $\hat r$ is in the production plane and defined as $\hat r = (\hat p_p - \cos \theta_t \hat k)/\sin \theta_t$, with $\theta_t$ the production angle in the centre-of-mass (c.m.) frame.
\item N-axis: $\hat n = \hat k \times \hat r$ is orthogonal to the production plane.
\end{itemize}
It is not our goal to provide a feasibility study for the observation of post-decay autodistillation, but to show that this effect is real and physically accessible at future colliders. We consider fully-polarized $e^+ e^-$ collisions at a c.m. energy of 1 TeV, generated with {\scshape MadGraph}~\cite{Alwall:2014hca}. Top pairs are produced in an almost pure state in polarized collisions. For central production $|\cos \theta_t| \leq 0.2$,  the $t \bar t$ state is
\begin{align}
& |\psi \rangle_{RL} = 0.32 \UU + 0.21 [\UD + \DU] + 0.90 \DD \,, \notag \\
& |\psi \rangle_{LR} = 0.92 \UU - 0.20 [\UD + \DU] + 0.25 \DD \,,
\end{align}
up to corrections at the permille level. (The subindices $RL$, $LR$ label the $e_R^+ e_L^-$ / $e_L^+ e_R^-$ initial state.) We consider the decay of the antiquark and the entanglement between the top quark and the $W^-$ boson, after tracing over the spins of the $b$ anti-quark (which has negligible influence because of its left-handed chirality, as previously discussed). As a measure of entanglement we use the negativity of the partial transpose on the $B$ subspace $\rho^{T_B}$,
\begin{equation}
N(\rho) = \frac{|| \rho^{T_B} || - 1}{2} \,,
\end{equation}
where $||A|| = \operatorname{tr} \sqrt{A A^\dagger} = \sum_i \sqrt{\lambda_i}$, where $\lambda_i$ are the (positive) eigenvalues of the matrix $AA^\dagger$.
We plot in Fig.~\ref{fig:Nvstheta} 
the negativity $N$ of the $tW^-$ pair as a function of $\theta$, for $\phi = 0,\pi$. For better comparison we also include the negativity of the $t \bar t$ pair.

For brevity we have shown here the entanglement amplification for fixed angles, but it is clear that the effect holds when one considers the integral over a sufficiently narrow region of the decay phase space. 
For example, for $e_R^+ e_L^-$ collisions the $tW$ entanglement is $N = 0.30$ in the region $\cos \theta \leq -0.9$~\cite{Aguilar-Saavedra:2024hwd}, i.e. larger than the initial $N=0.25$ in the $t\bar t$ system. We have also considered fully-polarized beams for simplicity, but the entanglement increase would also manifest with partial beam polarization with the expected parameters~\cite{Vauth:2016pgg}. For polarizations $P_{e^+} = 0.6$, $P_{e^-} = -0.9$, and since $\sigma(e_R^+ e_L^-) \simeq 2 \sigma(e_L^+ e_R^-)$, the contribution from $e_L^+ e_R^-$ is 150 times smaller than that of $e_R^+ e_L^-$, thus negligible. We also note that the $tW$ entanglement in the regions of interest is much larger than the one found in $pp$ collisions, and with sufficient statistics of $t \bar t$ pairs it would be measurable  using the same method discussed in Ref.~\cite{Aguilar-Saavedra:2023hss}.

\section{Discussion}

We have shown the remarkable fact that for a spin-entangled pair
of particles the entanglement of the system can spontaneously increase after the decay of one particle. 
This would not be possible for a purely unitary evolution; however, particle decay also entails the measurement of the final state products with definite momenta, which is a projection. And this contrasts with other phenomena that a subsystem can undergo in its interaction with the environment, which generally lead to decoherence. Autodistillation occurs with a certain probability, and thus can be considered as a kind of SLOCC \cite{Benenti}, as it is also standard distillation, 
i.e. the manufacturing of maximally-entangled copies of a system from a larger number of less-entangled copies \cite{Benenti}. Note however that distillation techniques require the use of ancillary qubits, which have to be prepared in a particular way that depends on a complete knowledge of the initial state. This step is absent in autodistillation, where everything takes place in a completely spontaneous way.

Finally, we have shown that this phenomenon would be experimentally observable in top pair production at a high-energy $e^+ e^-$ collider.  In other physical systems this effect might even be used to prepare states with large entanglement.

\section*{Acknowledgements}

This work has been supported by the Spanish Research Agency (Agencia Estatal de Investigaci\'on) through projects PID2019-110058GB-C21, PID2019-110058GB-C22, PID2022-142545NB-C21, PID2022-142545NB-C22 and CEX2020-001007-S funded by MCIN/AEI/10. 13039/501100011033, and by Funda\c{c}{\~a}o para a Ci{\^e}ncia e a Tecnologia (FCT, Portugal) through the project CERN/FIS-PAR/0019/2021.


\begin{thebibliography}{99}

\bibitem{Einstein:1935rr}
A.~Einstein, B.~Podolsky and N.~Rosen,
``Can quantum mechanical description of physical reality be considered complete?,''
Phys. Rev. \textbf{47}, 777-780 (1935)

\bibitem{sch1935}
E. Schrödinger, ``Discussion of Probability Relations between Separated Systems'',
Math. Proc. Cambridge Philos. Soc. 31, 555 (1935).


\bibitem{Aguilar-Saavedra:2023hss}
J.~A.~Aguilar-Saavedra,
``Post-decay quantum entanglement in top pair production,''
Phys. Rev. D \textbf{108}, no.7, 076025 (2023)
[arXiv:2307.06991 [hep-ph]].

\bibitem{Aguilar-Saavedra:2023lwb}
J.~A.~Aguilar-Saavedra,
``Decay of entangled fermion pairs with post-selection,''
Phys. Lett. B \textbf{848}, 138409 (2024)
[arXiv:2308.07412 [hep-ph]].


\bibitem{Bernabeu:2019gjs}
J.~Bernab\'eu and A.~Di Domenico,
``Can future observation of the living partner post-tag the past decayed state in entangled neutral K mesons?,''
Phys. Rev. D \textbf{105}, no.11, 116004 (2022)
[arXiv:1912.04798 [quant-ph]].

\bibitem{Bennett:2000fte}
C.~H.~Bennett, S.~Popescu, D.~Rohrlich, J.~A.~Smolin and A.~V.~Thapliyal,
``Exact and asymptotic measures of multipartite pure-state entanglement,''
Phys. Rev. A \textbf{63} (2000) no.1, 012307



\bibitem{Kane:1991bg}
G.~L.~Kane, G.~A.~Ladinsky and C.~P.~Yuan,
``Using the Top Quark for Testing Standard Model Polarization and CP Predictions,''
Phys. Rev. D \textbf{45}, 124-141 (1992)

\bibitem{Aguilar-Saavedra:2015yza}
J.~A.~Aguilar-Saavedra and J.~Bernab\'eu,
``Breaking down the entire W boson spin observables from its decay,''
Phys. Rev. D \textbf{93}, no.1, 011301 (2016)
[arXiv:1508.04592 [hep-ph]].

\bibitem{Rahaman:2021fcz}
R.~Rahaman and R.~K.~Singh,
``Breaking down the entire spectrum of spin correlations of a pair of particles involving fermions and gauge bosons,''
Nucl. Phys. B \textbf{984}, 115984 (2022)
[arXiv:2109.09345 [hep-ph]].

\bibitem{Ashby-Pickering:2022umy}
R.~Ashby-Pickering, A.~J.~Barr and A.~Wierzchucka,
``Quantum state tomography, entanglement detection and Bell violation prospects in weak decays of massive particles,''
JHEP \textbf{05}, 020 (2023)
[arXiv:2209.13990 [quant-ph]].

\bibitem{Bernal:2023jba}
A.~Bernal,
``Quantum tomography of helicity states for general scattering processes,''
Phys. Rev. D \textbf{109}, no.11, 116007 (2024)
[arXiv:2310.10838 [hep-ph]].

\bibitem{ATLAS:2023fsd}
ATLAS Collaboration,
``Observation of quantum entanglement in top-quark pairs using the ATLAS detector,''
[arXiv:2311.07288 [hep-ex]].

\bibitem{CMS:2024pts}
CMS Collaboration,
``Observation of quantum entanglement in top quark pair production in proton-proton collisions at $\sqrt{s}$ = 13 TeV,''
[arXiv:2406.03976 [hep-ex]].

\bibitem{Aguilar-Saavedra:2024hwd}
J.~A.~Aguilar-Saavedra,
``A closer look at post-decay $t \bar t$ entanglement,''
Phys. Rev. D \textbf{109}, no.9, 096027 (2024)
[arXiv:2401.10988 [hep-ph]].

\bibitem{Boudjema:2009fz}
F.~Boudjema and R.~K.~Singh,
``A Model independent spin analysis of fundamental particles using azimuthal asymmetries,''
JHEP \textbf{07}, 028 (2009)
[arXiv:0903.4705 [hep-ph]].



\bibitem{Afik:2020onf}
Y.~Afik and J.~R.~M.~de Nova,
``Entanglement and quantum tomography with top quarks at the LHC,''
Eur. Phys. J. Plus \textbf{136}, no.9, 907 (2021)
[arXiv:2003.02280 [quant-ph]].

\bibitem{Fabbrichesi:2021npl}
M.~Fabbrichesi, R.~Floreanini and G.~Panizzo,
``Testing Bell Inequalities at the LHC with Top-Quark Pairs,''
Phys. Rev. Lett. \textbf{127} (2021) no.16, 16
[arXiv:2102.11883 [hep-ph]].

\bibitem{Severi:2021cnj}
C.~Severi, C.~D.~Boschi, F.~Maltoni and M.~Sioli,
``Quantum tops at the LHC: from entanglement to Bell inequalities,''
Eur. Phys. J. C \textbf{82}, no.4, 285 (2022)
[arXiv:2110.10112 [hep-ph]].

\bibitem{Afik:2022kwm}
Y.~Afik and J.~R.~M.~de Nova,
``Quantum information with top quarks in QCD,''
Quantum \textbf{6}, 820 (2022)
[arXiv:2203.05582 [quant-ph]].

\bibitem{Aguilar-Saavedra:2022uye}
J.~A.~Aguilar-Saavedra and J.~A.~Casas,
``Improved tests of entanglement and Bell inequalities with LHC tops,''
Eur. Phys. J. C \textbf{82}, no.8, 666 (2022)
[arXiv:2205.00542 [hep-ph]].

\bibitem{Afik:2022dgh}
Y.~Afik and J.~R.~M.~de Nova,
``Quantum Discord and Steering in Top Quarks at the LHC,''
Phys. Rev. Lett. \textbf{130}, no.22, 221801 (2023)
[arXiv:2209.03969 [quant-ph]].

\bibitem{Dong:2023xiw}
Z.~Dong, D.~Gon\c{c}alves, K.~Kong and A.~Navarro,
``Entanglement and Bell inequalities with boosted $t \bar t$,''
Phys. Rev. D \textbf{109}, no.11, 115023 (2024)
[arXiv:2305.07075 [hep-ph]].
\bibitem{Han:2023fci}
T.~Han, M.~Low and T.~A.~Wu,
``Quantum Entanglement and Bell Inequality Violation in Semi-Leptonic Top Decays,''
[arXiv:2310.17696 [hep-ph]].


\bibitem{Bernreuther:2015yna}
W.~Bernreuther, D.~Heisler and Z.~G.~Si,
``A set of top quark spin correlation and polarization observables for the LHC: Standard Model predictions and new physics contributions,''
JHEP \textbf{12}, 026 (2015)
[arXiv:1508.05271 [hep-ph]].

\bibitem{Alwall:2014hca}
J.~Alwall, R.~Frederix, S.~Frixione, V.~Hirschi, F.~Maltoni, O.~Mattelaer, H.~S.~Shao, T.~Stelzer, P.~Torrielli and M.~Zaro,
``The automated computation of tree-level and next-to-leading order differential cross sections, and their matching to parton shower simulations,''
JHEP \textbf{07}, 079 (2014)
[arXiv:1405.0301 [hep-ph]].

\bibitem{Vauth:2016pgg}
A.~Vauth and J.~List,
``Beam Polarization at the ILC: Physics Case and Realization,''
Int. J. Mod. Phys. Conf. Ser. \textbf{40}, 1660003 (2016)

\bibitem{Benenti}
G. Benenti, G. Casati, D. Rosini and G. Strini,
``Principles of Quantum Computation and Information'' 
World Scientific, 2019.

\end{thebibliography}
\end{document}